\documentclass[
					aps, 							
					pre,							
					reprint,         
					showpacs, 					
					preprintnumbers,		
					amsmath,
					amssymb,					
					english
					]{revtex4-1}

\usepackage{dcolumn}	
\usepackage[english]{babel}
\usepackage{bm}

\begin{document}

\preprint{ES.R06/02-en24}

\title{Surface  in statistical ensembles}

\author{V.M. Zaskulnikov}
\homepage{http://www.zaskulnikov.ru}
\email[]{zaskulnikov@gmail.com }

\affiliation{Institute of Chemical Kinetics and Combustion, Institutskaya street 3, Novosibirsk, 630090, Russian Federation}

\date{\today}

\begin{abstract}

The present contribution deals with surface  terms appearing immediately in distributions and partition functions of statistical ensembles. It is shown that all ensembles under study, including ordinary canonical and grand canonical ensembles, involve surface terms. For  a canonical ensemble both surface and volume terms correspond to a closed system. For a grand canonical ensemble the volume term corresponds to an open system, while the surface one - to a closed one. Finally, for the recently introduced open statistical ensemble the specific feature of which is the consideration of some surrounding region both volume and surface terms correspond to an open system.

In conclusion, surface particles at solid/fluid boundary are interpreted as particles corresponding to number density oscillations near the surface; this completely agrees with the earlier introduced concept of surface tension at such a boundary.
 
\end{abstract}


\maketitle

\section{\label{sect:01}Introduction}

In paper \cite{zaskulnikov200911} an open statistical ensemble (OSE) is studied, and comparison with a grand canonical ensemble (GCE) is made. The main peculiarity of OSE is correct consideration of surface terms.

OSE approach enables one to examine surface terms directly on the basis of fundamental statistical relations rather than to introduce them additionally.

Unlike GCE, for OSE the analogs of compressibility equation for higher order and  `mixed' correlations hold rigorously.

The present paper shows that surface terms can be found even in a standard canonical ensemble (CE), and also correspond to the boundary between fluid and hard solid.

Thermodynamic analysis verifies the form of surface terms in CE and the expression for $\Omega$-potential of the CE as a closed system with the surface.

In papers \cite{zaskulnikov200911} and \cite{Bellemans1962} GCE surface terms are examined. Comparison between OSE and GCE made in \cite{zaskulnikov200911} demonstrates that surface terms are equal in value and opposite in sign. In this contribution we reveal the meaning of this fact related to hybrid nature of GCE.

Paper \cite{zaskulnikov200911} shows that unlike equilibrium two-phase system, for the case at hand the coefficient of surface tension depends on two variables, namely, pressure and temperature, and corresponds to the boundary between fluid and hard solid. This results in the non-zero number of surface particles for such a boundary. This paper establishes their physical meaning as particles lying within the limits of near-surface oscillations or other near-surface distortions corresponding to local deviation of density from the mean one.

\section{\label{sect:02}Primary definitions}

\subsection{\label{subsect:02a}Canonical ensemble}

Let us start with configuration integral 
\begin{equation}
Z_N = \int \limits_V \exp(-\beta U^N_{1...N}) d\bm{r}_1...d\bm{r}_N. 
\label{equ:001}
\end{equation}
Here $N$ is the number of particles in the system, $\beta = 1/k_BT$, $k_B$ is the Boltzmann constant, $T$ is the temperature, $U^N_{1...N}$ is the interaction energy of particles, $V$ is the system volume. Integration is performed over the coordinates of all particles of the ensemble $\bm{r}_1...\bm{r}_N$.

The probability density of finding a given space configuration of a particular set of particles \cite[p.181]{hillstatmeh1987} is given by the expression
\begin{equation}
P^{(k)}_{1...k} = \frac{1}{Z_N}\int \limits_V \exp(-\beta U^N_{1...N})d\bm{r}_{k+1}...d\bm{r}_N. 
\label{equ:002}
\end{equation}
Passing to distribution functions for arbitrary set of particles, we have  
\begin{equation}
\varrho^{(k)}_{C,1...k} = \frac{N!}{(N-k)!}P^{(k)}_{1...k}, 
\label{equ:003}
\end{equation}
where $\varrho^{(k)}_{C,1...k}$ defines the probability density of finding a given configuration of $k$ arbitrary particles for the canonical ensemble.

\subsection{\label{subsect:02b}Grand canonical ensemble}

Average equality (\ref{equ:003}) over fluctuations of the number of particles, i.e., apply the operation $\sum_{N=0}^\infty P_N$ to both its sides. Here
\begin{equation}
P_N = \frac{z^N Z_N}{N!\Xi_V} 
\label{equ:004}
\end{equation}
is the probability for GCE to have a certain number $N$ of particles within the volume $V$.  $z$ is the activity
\begin{equation}
z = \frac{e^{\mu/k_BT}}{\Lambda^3}, 
\label{equ:005}
\end{equation}
where $\mu$ is the chemical potential, $\Lambda = h/\sqrt[]{2 \pi mk_BT}$, $h$ is the Planck constant, $m$ is the particle mass, and $\Xi_V$ is a large partition function of the system of the volume $V$
\begin{equation}
\Xi_V = 1 + \sum_{N=1}^\infty \frac{z^N Z_N}{N!}. 
\label{equ:006}
\end{equation}
We obtain
\begin{equation}
\varrho^{(k)}_{G,1...k} = \sum_{N=k}^\infty \varrho^{(k)}_{C,1...k} P_N,
\label{equ:007}
\end{equation}
or
\begin{eqnarray}
\varrho^{(k)}_{G,1...k} &=& \frac{z^k}{\Xi_V}\exp(-\beta U^{k}_{1...k}) +  \frac{z^k}{\Xi_V}\sum_{N=1}^\infty \frac{z^N}{N!} \label{equ:008} \\
&& \times \int \limits_V    \exp(-\beta U^{N+k}_{1...N+k})d\bm{r}_{k+1}...d\bm{r}_{k+N}.  \nonumber
\end{eqnarray}
$\varrho^{(k)}_{G,1...k}$ specify the probability density of finding certain configuration of arbitrary particles for GCE. For ideal gas $\varrho^{(k)}_{G,1...k} = \varrho^{k}$, where $\varrho = \overline{N}/V$ is the density of particles.

\subsection{\label{subsect:02c}Open statistical ensemble}

OSE functions corresponding to $\varrho^{(k)}_{G,1...k}$ are defined as
\begin{equation}
\varrho^{(k)}_{1...k}(z)=\lim_{V\rightarrow \infty}\varrho^{(k)}_{G,1...k}(z),
\label{equ:009}
\end{equation}
on condition that the region containing the coordinates of particles remains finite, and intensive parameters of the medium are kept unchanged. In practice, this means that we can use the definition of GCE correlation functions rather far away from the boundaries of the ensemble volume.

In contrast to GCE correlation functions, the expansion $\varrho^{(k)}_{1...k}$ into a series in powers of activity involves no window functions, and integration  in the integrals of the expansion coefficients is performed over the infinite space \cite{zaskulnikov200911}.

\subsection{\label{subsect:02d}Localized correlation functions}

Application of  logarithmic generating function gives rise to Ursell functions or localized correlation functions \cite[]{ursell1927}, \cite[]{percus1964}.

These functions have the property of locality: they decay rapidly when any groups of particles move apart including the case with one particle moving away from others.  We give several first functions for a homogeneous medium.
\begin{eqnarray}
{\cal F}^{(1)}_{1}\mkern 15mu & = & \varrho\label{equ:010}  \\
{\cal F}^{(2)}_{1,2} \mkern 16mu& = & \varrho^{(2)}_{1,2} -  \varrho^2 \nonumber \\
{\cal F}^{(3)}_{1,2,3}\mkern 10mu & = & \varrho^{(3)}_{1,2,3} -  ( \varrho^{(2)}_{1,2} + \varrho^{(2)}_{2,3} + \varrho^{(2)}_{1,3})\varrho  + 2\varrho^{3} \nonumber \\
{\cal F}^{(4)}_{1,2,3,4} & = & \varrho^{(4)}_{1,2,3,4} -  (\varrho^{(3)}_{1,2,3} + \varrho^{(3)}_{2,3,4} + \varrho^{(3)}_{1,3,4} + \varrho^{(3)}_{1,2,4})\varrho\nonumber \\
& - & (\varrho^{(2)}_{1,2}\varrho^{(2)}_{3,4}  +  \varrho^{(2)}_{2,3}\varrho^{(2)}_{1,4} + \varrho^{(2)}_{1,3}\varrho^{(2)}_{2,4})+ 2(\varrho^{(2)}_{1,2} + \varrho^{(2)}_{2,3}\nonumber \\
& + & \varrho^{(2)}_{1,3}  +  \varrho^{(2)}_{1,4}  +  \varrho^{(2)}_{2,4} + \varrho^{(2)}_{3,4})\varrho^2 - 6\varrho^{4}
\nonumber \\
\dotso \nonumber 
\end{eqnarray}
Obviously, these functions can be formed for any ensemble, and the equality valid for arbitrary $k \geq 1$ may serve as their definition
\begin{eqnarray}
{\cal F}^{(k)}_{1...k} &=& \sum_{\{\bm{n}\}}(-1)^{l-1}(l-1)!\prod_{\alpha = 1}^l \varrho^{(k_\alpha)}(\{\bm{n}_\alpha\}), \nonumber \\
1 & \leq & k_\alpha \leq k, ~~~~~~~ \sum_{\alpha = 1}^l k_\alpha = k, \label{equ:011}
\end{eqnarray}
where $\{\bm{n}\}$ denotes some partitioning of a given set of $k$ particles with the coordinates $\bm{r}_1,...\bm{r}_k$ into disjoint groups $\{\bm{n}_\alpha\}$, $l$ is the number of groups of a specific partitioning, $k_\alpha$ is the size of the group with the number $\alpha$, summation is taken over all possible partitionings.

The inverse relation is valid
\begin{equation}
\varrho^{(k)}_{1...k} = \sum_{\{\bm{n}\}}\prod_{\alpha = 1}^l {\cal F}^{(k_\alpha)}(\{\bm{n}_\alpha\}),
\label{equ:012}
\end{equation}
with  notation the same as in (\ref{equ:011}) \cite[Chapter 9]{MayerGeppert1977}.

\subsection{\label{subsect:02e}Ursell factors}

Ursell factors ${\cal U}^{(k)}_{1...k}$ are also called cluster functions. These are the functions that  appear in the familiar pressure expansion in powers of activity (\ref{equ:037}) \cite[p.129]{hillstatmeh1987}, \cite[p.232]{landaulifshitz1985}.

They may be defined by the equality
\begin{eqnarray}
{\cal U}^{(k)}_{1...k} &=& \sum_{\{\bm{n}\}}(-1)^{l-1}(l-1)!\prod_{\alpha = 1}^l \exp(-\beta U^{k_\alpha}(\{\bm{n}_\alpha\})),  \nonumber \\
1 & \leq & ~ k_\alpha \leq k, ~~~ \sum_{\alpha = 1}^l k_\alpha = k, ~~~ \exp(-\beta U^{1}) = 1,
\label{equ:013}
\end{eqnarray}
where designations are the same as in (\ref{equ:011}), and the meaning of the condition $\exp(-\beta U^{1}) = 1$ is evident.

As an example we give several first ${\cal U}^{(k)}_{1...k}$
\begin{eqnarray}
{\cal U}^{(1)}_{1}\mkern 9mu &=& 1 \label{equ:014} \\
{\cal U}^{(2)}_{1,2} \mkern 8mu &=& \exp(-\beta U^{2}_{1,2}) - 1 \nonumber \\
{\cal U}^{(3)}_{1,2,3} &=& \exp(-\beta U^{3}_{1,2,3}) - \exp(-\beta U^{2}_{1,2}) \nonumber \\
 &-& \exp(-\beta U^{2}_{1,3}) - \exp(-\beta U^{2}_{2,3}) + 2 \nonumber \\
\dotso \nonumber
\end{eqnarray}
Ursell  factors also decay rapidly  when any group of particles, including  a single particle, moves away.

\section{\label{sect:03}Canonical ensemble}

First we shall describe the mechanism of calculating the first surface terms of the series in activity in this ensemble, then general proof will be given.

\subsection{\label{subsect:03a}The series in the group size}

Let us employ expressions (\ref{equ:011}) and (\ref{equ:013}) that yield analog (\ref{equ:012}) - the Bolzmann factor expansion in Ursell factors
\begin{eqnarray}
\exp(-\beta U^{k}_{1...k}) &=& \sum_{\{\bm{n}\}}\prod_{\alpha = 1}^l {\cal U}^{(k_\alpha)}(\{\bm{n}_\alpha\}), \nonumber \\
1 \leq k_\alpha & \leq & k, ~~~~~~~ \sum_{\alpha = 1}^l k_\alpha = k. \label{equ:015}
\end{eqnarray}
As before, designations are the same as in (\ref{equ:011}).

As far as we know, equation (\ref{equ:015}) was first derived in \cite[]{ursell1927}, and was used for the terms of configuration integral logarithm expansion in powers of activity.

So far it has probably remained unrealized that expansion (\ref{equ:015}) may be treated as a series in the parameter which is the total number of particles in nonsingle groups (the sum of sizes of the partition groups except single particle groups). In view of
\begin{equation}
{\cal U}^{(1)}_1 = 1,
\label{equ:016}
\end{equation}
and changing the indices, we rewrite equation (\ref{equ:015}) as
\begin{eqnarray}
\exp(-\beta U^{N}_{1...N}) &=& 1 + \sum_{k=2}^N \sum_{\{\bm{n}\}}\prod_{\alpha = 1}^l {\cal U}^{(k_\alpha)}(\{\bm{n}_\alpha\}), \nonumber \\
2 \leq k_\alpha & \leq & k, ~~~~~~~ \sum_{\alpha = 1}^l k_\alpha = k, \label{equ:017}
\end{eqnarray}
with  notation also the same as in (\ref{equ:011}) except that partitionings are made over nonsingle groups.

It is easily seen that, unlike (\ref{equ:015}), (\ref{equ:017}) explicitly demonstrates the absense of single particle groups (isolated particles).

Thus the principal idea of this section is to apply (\ref{equ:017}) immediately to the entire Boltzmann factor included in $Z_N$, i.e., to use  the partitioning of the whole combination of $N$ particles contained in the closed system into groups.

If expanded, (\ref{equ:017}) has the form of a series in the total size of the groups
\begin{widetext}
\begin{equation}
\exp(-\beta U^{N}_{1...N}) = 1 + \sum^{}_{1\leq i<j \leq N} {\cal U}^{(2)}_{i,j}  +  \sum^{}_{1\leq i<j<k \leq N} {\cal U}^{(3)}_{i,j,k}  + \sum^{}_{1\leq i<j<k<l \leq N} \left \{{\cal U}^{(4)}_{i,j,k,l} + {\cal U}^{(2)}_{i,j} {\cal U}^{(2)}_{k,l} + {\cal U}^{(2)}_{i,k} {\cal U}^{(2)}_{j,l} + {\cal U}^{(2)}_{i,l} {\cal U}^{(2)}_{k,j} \right \} + \dotso  
\label{equ:018}
\end{equation} 
\end{widetext} 
Substitute (\ref{equ:017}) in (\ref{equ:001}). Making integration over the coordinates of free particles and grouping identical terms, we pass to a series for configuration integral
\begin{eqnarray}
\frac{Z_N}{V^N} &=&1 + \sum_{k=2}^N \frac{N!}{(N-k)! V^k} \sum_{\{\bm{m}\}} \prod_{j = 2}^k \frac{c_j^{m_j}}{m_j!}, \nonumber\\
& &  \sum_{j = 2}^k j m_j = k. \label{equ:019}
\end{eqnarray}
Here in the internal sum we go from the sum over partitionings to the sum over the partitioning topology: $\{\bm{m}\}$ is some partitioning of the group of $k$ particles into subgroups of the size $j$ irrespective of the numbers of particles, $m_j$ is the quantity of subgroups of the size $j$, summation is taken over all possible partitionings, and $c_j$ has the form
\begin{equation}
c_j = \frac{1}{j!}\int \limits_V{ {\cal U}^{(j)}_{1...j} d\bm{r}_1... d\bm{r}_j}.
\label{equ:020}
\end{equation}
Expression (\ref{equ:019}) is similar to the expression given in \cite[p.128]{hillstatmeh1987}, but with single particle groups explicitly eliminated from it, and with traditional expression $Vb_j$ replaced by $c_j$ the meaning of which will become clear later on. In the expanded form
\begin{widetext}
\begin{equation}
\frac{Z_N}{V^N} = 1~ +~ \frac{1}{V^2} \binom{N}{2} \int\limits_V{{\cal U}^{(2)}_{1,2}} d\bm{r}_1 d\bm{r}_2~  +~  \frac{1}{V^3} \binom{N}{3} \int\limits_V{{\cal U}^{(3)}_{1,2,3}} d\bm{r}_1 d\bm{r}_2 d\bm{r}_3~  +~ \frac{1}{V^4} \binom{N}{4} \int\limits_V{  \left ({\cal U}^{(4)}_{1...4} + 3 {\cal U}^{(2)}_{1,2} {\cal U}^{(2)}_{3,4} \right )d\bm{r}_1... d\bm{r}_4} + \dotso  
\label{equ:021}
\end{equation} 
\end{widetext}
Generally speaking, this series (in density) is not a series in a small parameter, however, it is structurally similar to exponent expansion, so we can take a logarithm of it and see that the obtained series corresponds to correct volume and surface terms. In other words, when taking a logarithm , for example, of the term with $k=2$, we actually make allowance for the products of all pair groups: double, triple, etc., entering in  sum (\ref{equ:021}). The analysis presented solves the problems with the series for configuration integral \cite[p.227]{landaulifshitz1985}.

\subsection{\label{subsect:03b}Separating out surface terms}

Introduce window functions of two types
\begin{equation}
\psi^V(\boldsymbol{r}_i) = \psi^V_i = 
	 \left\{ 
			\begin{array}{ll} 
         1 & (\boldsymbol{r}_i \in V)\\   
         0 & (\boldsymbol{r}_i \notin V).
     	\end{array}  
		\right.
		\label{equ:022}
\end{equation}
and
\begin{equation}
\chi^V_i = 1 - \psi^V_i = 
	 \left\{ 
			\begin{array}{ll} 
         0 & (\bm{r}_i \in V)\\   
         1 & (\bm{r}_i \notin V).
     	\end{array}  
		\right.
		\label{equ:023}
\end{equation}
Application of $\psi^V_i$ and $\chi^V_i$ implies that integration is performed over the infinite space, unless otherwise specified.

Rewrite expression (\ref{equ:020}) with the use of these functions. This gives
\begin{equation}
c_j = \frac{1}{j!}\int {\left [ \prod_{k = 1}^j \psi^V_k \right ] {\cal U}^{(j)}_{1...j} d\bm{r}_1... d\bm{r}_j}.
\label{equ:024}
\end{equation}
The procedure of separating out surface terms \cite{zaskulnikov200911} yields
\begin{equation}
c_j = Vb_j + Aa_j,
\label{equ:025}
\end{equation}
where $A$ is the surface surrounding the system, $Vb_j$ corresponds to traditional definition \cite[p.129]{hillstatmeh1987}
\begin{equation}
Vb_j = \frac{1}{j!}\int  {\psi^V_1 {\cal U}^{(j)}_{1...j} d\bm{r}_1... d\bm{r}_j},
\label{equ:026}
\end{equation}
and
\begin{equation}
Aa_j =  \frac{1}{j!}\int {\psi^V_1 \left [ \prod_{k = 2}^j \psi^V_k - 1\right ] {\cal U}^{(j)}_{1...j} d\bm{r}_1... d\bm{r}_j},
\label{equ:027}
\end{equation}
or, with $\psi^V_k$ replaced by $1-\chi^V_k$ between the brackets,
\begin{equation}
Aa_j =  \frac{1}{j!}\int {\psi^V_1 \left [ \prod_{k = 2}^j (1 - \chi^V_k) - 1\right ] {\cal U}^{(j)}_{1...j} d\bm{r}_1... d\bm{r}_j}.
\label{equ:028}
\end{equation}
After the product expansion the integrands in (\ref{equ:028}) always involve at least one pair of particles on different sides of the system boundary, thus providing the proportionality of this expression to its surface \cite{zaskulnikov200911}.

Below we show that the presence of surface terms in (\ref{equ:025}) is not accidental, and that boundary mathematical effects coincide closely with physical ones.

In pure form for $a_j$ and $b_j$ we have 
\begin{equation}
b_j = \frac{1}{j!}\int  {{\cal U}^{(j)}_{1...j} d\bm{r}_2... d\bm{r}_j},
\label{equ:029}
\end{equation}
\begin{equation}
a_j =  \frac{1}{j!}\int {\psi^V_1 \left [ \prod_{k = 2}^j (1 - \chi^V_k) - 1\right ] {\cal U}^{(j)}_{1...j} dx_1 d\bm{r}_2... d\bm{r}_j},
\label{equ:030}
\end{equation}
where $x_1$ is the coordinate perpendicular to the system surface, and the  axis direction corresponds to positive increment.

So the partition function of the canonical ensemble involves surface terms. Their exact form may be obtained by substituting (\ref{equ:025}) in (\ref{equ:019}), and, if necessary, taking a logarithm of the series.

The first terms of the expansion are
\begin{widetext}
\begin{equation}
\frac{Z_N}{V^N} = 1~ +~  \frac{N!}{(N-2)!} \frac{(Vb_2 + Aa_2)}{V^2}~  +~ \frac{N!}{(N-3)!} \frac{(Vb_3 + Aa_3)}{V^3}~  +~ \frac{N!}{(N-4)!} \frac{ \left [ (Vb_4 + Aa_4) + \frac{1}{2} (Vb_2 + Aa_2)^2 \right ]}{V^4} + \dotso  
\label{equ:031}
\end{equation} 
\end{widetext}
Taking a logarithm and in view of
\begin{eqnarray}
F_N &=&N k_B T\ln (N \Lambda^3)- N k_B T - k_B T \ln Z_N; \nonumber\\
& &  \Omega_N = F_N - \mu N, \label{equ:032}
\end{eqnarray}
where $F_N$ is free energy and $\Omega_N$ is omega potential of the closed system, we get
\begin{eqnarray}
\Omega_N = -V k_B T(z &+& z^2 b_2 + z^3 b_3 + \dotso) \nonumber\\
& &   -A k_B T(z^2 a_2 + z^3 a_3 + \dotso). \label{equ:033}
\end{eqnarray}
Here we go from the series in powers of density to the series in powers of activity using the known expansion
\begin{equation}
\varrho(z) = z + z \sum_{n=1}^\infty (n+1) z^n b_{n+1}.
\label{equ:034}
\end{equation}
In the next section we shall establish the general form of the $\Omega$-potential of the canonical ensemble, however, first it is necessary to consider grand canonical ensemble.

\section{\label{sect:04}Grand canonical ensemble}

In \cite{zaskulnikov200911} the procedure of separating out volume and surface terms similar to that described above was applied to GCE partition function
\begin{equation}
\ln{\Xi_V} =  \sum_{t=1}^\infty \frac{z^t}{t!} \int \left [ \prod_{i = 1}^{t} \psi^V_i \right ]  {\cal U}^{(t)}_{1...t} d\boldsymbol{r}_1...d\boldsymbol{r}_t.
\label{equ:035}  
\end{equation}
It has been shown that
\begin{equation}
\Xi_V = \exp{\beta [ VP(z,T)  - A\sigma (z,T) ]},
\label{equ:036}
\end{equation}
where
\begin{equation}
P(z,T) =zk_BT  + k_BT \sum_{k=2}^\infty \frac{z^k}{k!} \int {\cal U}^{(k)}_{1...k} d\boldsymbol{r}_2...d\boldsymbol{r}_k,
\label{equ:037} 
\end{equation}
is the well-known expansion for pressure, and for surface tension coefficient we have the expression
\begin{eqnarray}
\sigma (z,T) &=& k_BT\sum_{t=2}^\infty \frac{z^t}{t!} \label{equ:038} \\
&\times &  \int \psi^V_1 \left [1 - \prod_{i = 2}^{t} (1 - \chi^V_i) \right ] {\cal U}^{(t)}_{1...t} dx_1 d\bm{r}_2 ...d\bm{r}_t.\nonumber
\end{eqnarray}
Equation (\ref{equ:036}) defines usual relation between partition function and GCE $\Omega$-potential
\begin{equation}
-k_B T\ln{\Xi_V} =  \Omega = -P V + \sigma A,  \label{equ:039}  
\end{equation}
but  the $\Omega$-potential itself has an unusual form. First, it corresponds to the system with a surface, second, now the surface tension coefficient depends on two variables. However, the first point is not surprising, since GCE is formed as a set of closed systems. Naturally, if we consider the systems each of which has a surface, this also refers to a combination of these systems.

The surface term sign corresponds exactly to a closed system, and contradicts thermodynamic expression for fluctuation probability in an open one \cite{zaskulnikov200911}.

So GCE is an open system from the standpoint of volume properties, since it describes adequately the fluctuation of the total number of particles. 

However, from the point of view of surface properties, it is a closed system.  This is primarily supported by the fact that for GCE the mean number of particles is equal to the sum of volume and surface ones \cite{zaskulnikov200911}. Probably, it is not coincidental that the author of paper \cite{Bellemans1962} has come to the conclusion that GCE describes a "drop".

Return to expression (\ref{equ:019}). As mentioned above, it is actually identical to the expression derived in \cite[p.128]{hillstatmeh1987} with the substitution
\begin{equation}
V b_j \rightarrow c_j.
\label{equ:040}
\end{equation}
With saddle-point technique it has been proved \cite[p.150]{hillstatmeh1987} that for the canonical ensemble we have the following expression for the $\Omega$-potential accurate to the terms logarithmic in the number of particles
\begin{equation}
- k_B T\sum_{j \geq 1}V b_j z^j.
\label{equ:041}
\end{equation}
Proceeding as in \cite[p.150]{hillstatmeh1987} with the use of substitution (40), we obtain
\begin{equation}
\Omega_N = - k_B T\sum_{j \geq 1} c_j z^j = -V P(z,T) + A \sigma (z,T),
\label{equ:042}
\end{equation}
where $P(z,T)$ and $\sigma (z,T)$ are defined by expressions (\ref{equ:037}) and (\ref{equ:038}), respectively.

Thus series (\ref{equ:033}) is nothing else but the first terms of the expansion in powers of activity of expression (\ref{equ:042}).

\section{\label{sect:05}Open statistical ensemble}

First we give some principal results derived in \cite{zaskulnikov200911} for OSE. Unlike (\ref{equ:035}), (\ref{equ:036}), for the partition function $\Upsilon_v$ of this  ensemble, we have the expressions
\begin{equation}
\ln{\Upsilon_V} = \sum_{t=1}^\infty \frac{z^t}{t!} \int \left [1 - \prod_{i = 1}^{t} \chi^V_i \right ] {\cal U}^{(t)}_{1...t} d\bm{r}_1...d\bm{r}_t
\label{equ:043} 
\end{equation}
and
\begin{equation}
\Upsilon_V = \exp{\beta [ VP(z,T)  + A\sigma (z,T) ]}.
\label{equ:044}
\end{equation}  
In contrast to GCE, in OSE the mean number of particles contains no surface terms, thus the $\Omega$-potential of OSE - $\Omega_V$ - corresponds to a true open system
\begin{equation}
\Omega_V = -P V.  
\label{equ:045}  
\end{equation}  
Naturally,
\begin{equation}
-k_B T\ln{\Upsilon_V} =  \Omega_V - \sigma A.  \label{equ:046}  
\end{equation}
Equality (\ref{equ:046}) seems logical, since the system thermodynamics is not to be determined solely by the probability of large deviations which is related to $p_0$, and, therefore, to $\Upsilon_V$. In other words, in (\ref{equ:046}) we see that, unlike (\ref{equ:039}), the partition function of the system is determined both by thermodynamic potential ($\Omega$) related to the mean values, and by the surface term related to large deviations.

\section{\label{sect:06}Surface particles}

So, since, according to (\ref{equ:038}), $\sigma$ depends on two variables, we obtain the non-zero value of the number of surface particles
\begin{equation}
N_S = - A \left ( \frac{\partial \sigma}{\partial \mu} \right )_T = - \beta A z \left ( \frac{\partial \sigma}{\partial z} \right )_T,
\label{equ:047}
\end{equation}
or, using (\ref{equ:038}),  
\begin{equation}
N_S = \sum_{t=2}^\infty \frac{z^t}{(t-1)!} \int \psi^V_1 \left [\prod_{i = 2}^{t} \psi^V_i - 1 \right ] {\cal U}^{(t)}_{1...t} d\bm{r}_1 ...d\bm{r}_t.
\label{equ:048}
\end{equation}
Equality (\ref{equ:048}) is almost apparent if we consider the familiar formula for density expansion into a series in powers of activity for GCE
\begin{equation}
\varrho_G(\bm{r}_{1},z) = z + z\sum_{n=1}^\infty \frac{z^n}{n!} \int \left [ \prod_{i = 2}^{n+1} \psi^V_i \right ]{\cal U}^{(n+1)}_{1...n+1} d\bm{r}_{2}...d\bm{r}_{n+1}.
\label{equ:049}
\end{equation}
and similar value for OSE
\begin{equation}
\varrho(z) = z + z\sum_{n=1}^\infty \frac{z^n}{n!} \int {\cal U}^{(n+1)}_{1...n+1} d\bm{r}_{2}...d\bm{r}_{n+1},
\label{equ:050}
\end{equation}
corresponding to (\ref{equ:037}). 

The number of surface particles in this case is identified with the value
\begin{equation}
N_S = \int \limits_V { \left [\varrho_G(\bm{r},z) - \varrho(z) \right ]d \bm{r}},
\label{equ:051}
\end{equation}
and surface particle density $\varrho_S = N_S/A$ is given by beautiful formula
\begin{equation}
\varrho_S = \int \limits_{L_t} { \left [\varrho_G(x,z) - \varrho(z) \right ] d x},
\label{equ:052}
\end{equation}
where $x$, as before - the coordinate perpendicular to the surface, and $L_t$ - the transition region near the surface.

In (\ref{equ:051}), (\ref{equ:052}) we employ the expression for boundary density of particles at the outer boundary of GCE. However, it is easily seen that the conclusion remains unchanged for the case of hard solid placed in the system. Moreover, result is the same for similar consideration performed not on the factor level but on the level of correlation functions \cite{zaskulnikov200911}. This points to the fact that the result does not depend on the value of the density parameter of the system.

Thus, the surface tension corresponds exactly to the number density oscillations (deviations) near the surface. We emphasize again that we are still talking about the boundary of the fluid and hard solid.

\end{document}